\documentclass[a4paper,11pt]{article}
\usepackage{pos}

\renewcommand{\logo}{\relax}

\usepackage{amsmath}
\usepackage{subcaption}
\usepackage{epsfig}
\usepackage{color} 
\usepackage{soul}
 \usepackage{multirow}
 \usepackage{enumitem}
\usepackage{amsfonts}
\usepackage[output-decimal-marker={.}]{siunitx}

\title{Charmonium-like states with $J^{P}=1^{+}$ and isospin 1 \note[\dag]{MITP-22-100}}

\author*[a]{Mitja Sadl}
\author[b]{Sara Collins}
\author[c,d]{M. Padmanath}
\author[a,e]{Sasa Prelovsek}

\affiliation[a]{Faculty of Mathematics and Physics, University of Ljubljana, 1000 Ljubljana, Slovenia}

\affiliation[b]{Institut für Theoretische Physik, Universität Regensburg, 93040 Regensburg, Germany}

\affiliation[c]{Helmholtz Institut Mainz, 55128 Mainz, Germany}

\affiliation[d]{GSI Helmholtzzentrum für Schwerionenforschung, 64291 Darmstadt, Germany}

\affiliation[e]{Jozef Stefan Institute, 1000 Ljubljana, Slovenia}

\emailAdd{mitja.sadl@fmf.uni-lj.si}
\emailAdd{sara.collins@ur.de}
\emailAdd{papppan@gmail.com}
\emailAdd{sasa.prelovsek@ijs.si}

\abstract{Many mesons with properties incompatible with a $\bar cc$ structure have already been discovered, e.g. the $Z_c$ mesons with isospin 1. We investigate the spectrum of exotic charmonium-like mesons using lattice QCD. The focus is on $\bar cc \bar qq$ states with $J^{PC}=1^{+\pm}$ and isospin 1. This is the first study of four-quark states with these quantum numbers,  a non-zero total momentum and two different lattice volumes. We extract the energy levels and determine the scattering length for $D\bar D^*$ scattering close to the threshold  using Lüscher's formalism. Our preliminary results show that the energy shifts for eigenstates dominated by $D\bar{D}^*$ are very small in the $1^{++}$ channel and consistent with zero in the $1^{+-}$ channel.}

\FullConference{%
  The 39th International Symposium on Lattice Field Theory (Lattice2022),\\
  8-13 August, 2022 \\
  Bonn, Germany 
}

\begin{document}
\maketitle

\section{Introduction}\label{sec:intro}

The first signal of a nonconventional meson was the discovery of the $\chi_{c1}(3872)$ by Belle in 2003 \cite{X(3872)_exp_1}. Its quantum numbers $I(J^{PC})=0(1^{++})$ are compatible with a naive $\bar{c}c$ structure, however, its mass and decay properties point to a more complex nature. The clearest evidence that a resonance containing  $\overline{c}c$ cannot be described as a simple quark-antiquark state is when it decays into a charged final state. The first such charged structure in the charmonium sector was discovered in 2013 when the BESIII and Belle collaborations observed the $Z_c^+(3900)$ in the $\pi^+\pi^-J/\psi$ invariant mass spectrum \cite{Zc(3900)_exp_1,Zc(3900)_exp_2}. This observation was confirmed by CLEO-c \cite{Zc(3900)_exp_4}. The quark content of the $Z^+_c(3900)$  is $\bar{c}c\bar{d}u$ ($I_z=1$). Its neutral partner $Z_c^0(3900)$ has also been seen \cite{Zc(3900)_exp_4,Zc(3900)_exp_3}. The invariant mass of the $Z_c(3900)$ lies slightly above the $D\bar{D}^*$ threshold suggesting that it could be observed in the decay channel $(D\bar{D}^*)_{I=1}$. This was confirmed by the BESIII collaboration \cite{Zc(3900)_exp_5,Zc(3900)_exp_6}. The current consensus is that $Z_c(3900)$ is a $1(1^{+-})$ state with mass $M=3887.1\pm\SI{2,6}{MeV}$ and width $\Gamma=28.4\pm\SI{2,6}{MeV}$ \cite{Workman:2022ynf}. Higher up in the spectrum,  the $Z_c(4200)$ \cite{Zc(4200)_exp} and the $Z_c(4430)$ \cite{Zc(4430)_exp_1,Zc(4430)_exp_2,Zc(4430)_exp_3,Zc(4430)_exp_4},  have also been seen.

Different binding mechanisms have been suggested for the $Z_c(3900)$: it could be a hadronic molecule, have a compact tetraquark structure or result from a simple kinematic effect linked to the opening of meson-meson thresholds. Many studies involving different effective field theory approaches have been performed. Combining local hidden gauge and heavy quark spin symmetry, ref. \cite{PhysRevD.90.016003}  finds that the exchange of heavy vector mesons gives the most significant contribution. The resulting scattering amplitude contains information about a state with a mass between 3869 and $\SI{3875}{MeV}$ and a decay width of around $\SI{40}{MeV}$. Another work  \cite{ALBALADEJO2016337}, which studies the invariant mass distribution of the $J/\psi \pi$ and $D\bar{D}^*$ channels suggests that the $Z_c(3900)$ signal may originate from a resonance or a virtual state, depending on whether the $D\bar{D}$ $s$-wave interaction employed is energy dependent or independent, respectively. If the peak is produced by a virtual state, it must have a hadronic molecular nature. The authors of \cite{He2018} come to similar conclusions. An analysis of the S-matrix poles in the framework of the constituent quark model involving coupled channels \cite{Ortega2019} connects the $Z_c(3900)$ signal with the presence of a virtual state that can be seen as a $D\bar{D}^*$ threshold cusp, i.e. a feature caused by the opening of a new threshold. This analysis is consistent with the interpretation that the diagonal interaction between the $D\bar{D}^{*}$ is too suppressed to develop resonances and that the interaction between different channels is responsible for a peak in the $D\bar{D}^*$, $J/\psi\pi$ invariant mass distributions.

Several lattice studies of the $Z_c(3900)$ have been performed so far: two works by the HAL QCD collaboration \cite{PhysRevLett.117.242001,Ikeda_2018} suggest the importance of cross-channel interaction, which is consistent with the conclusions of ref. \cite{Ortega2019}. However, works which employ Lüscher's formalism have not been able to confirm a narrow resonance-like peak close to the threshold. This includes \cite{Zc_Sasa_1,Zc_Sasa_2,Zc_CLQCD,Lee:2014uta,Cheung2017} and the more recent coupled channel analysis of \cite{CLQCD:2019npr}. In particular, no additional eigenstates are found and the energy shifts with respect to the non-interacting levels turn out to be insignificant. Comparing results from both methods is difficult since the HAL QCD approach does not provide information on the energy shifts. 

While charmonium-like states with $1(1^{+-})$ have been discovered in experiment, no states with $1(1^{++})$ and quark content $\bar{c}c\bar{d}u$ have been observed. Such a state would be an isospin partner of the $\chi_{c1}(3872)$. Two lattice QCD studies \cite{TOP_1_5,PRD_2015}, which find the state $\chi_{c1}(3872)$ slightly below the $D\bar{D}^*$ threshold, also do not see any new candidates in this spectrum.

In this proceedings, we report on a lattice study of charmonium-like states with quantum numbers $1(1^{+\pm})$. We employ meson-meson interpolating operators that are projected on to two different total momenta. The corresponding two-point correlation functions are calculated on two lattices with different spatial extents. The extraction of the energy levels is challenging since we are interested in the region near the $D\bar{D}^*$ threshold, which lies above several other meson-meson thresholds, e.g., $J/\psi\pi$ and $\eta_c\rho$ in the $1(1^{+-})$ channel and $J/\psi\rho$ in the $1(1^{++})$ channel.

\section{Lattice details}\label{sec:details}

We employ two ensembles of gauge field configurations with $N_{\textrm{f}}=2+1$ non-perturbatively $\mathcal{O}(a)$ improved Wilson dynamical fermions, a lattice spacing $a=0.08636(98)(40)\,$fm and a pion mass $m_\pi = \SI{280(3)}{MeV}$. The ensembles are provided by the Coordinated Lattice Simulations consortium \cite{Bruno2015,PhysRevD.94.074501}. The spatial volumes are $N_L^3=24^3$ and $N_L^3=32^3$, where we utilise $255$ and $492$ configurations, respectively \cite{PhysRevD.95.074504}. Open boundary conditions in time are imposed \cite{Luscher:2012av} and the sources of the correlation functions are located in the bulk away from the boundary. The study is performed for a charm quark mass which is slightly larger than the physical quark mass \cite{sasa_charmonia_vector}.

\section{Interpolating operators}\label{sec:interpolators}

The finite-volume energies are determined from the correlation matrices
\begin{equation}C_{ij} (t) = \langle O_i(t_{\textrm{src}}+t)O^{\dagger}_j(t_{\textrm{src}}) \rangle\>,
\label{E0}
\end{equation}
where $O_i$ ($O_j^{\dagger}$) is an interpolator that annihilates (creates) a state with certain quantum numbers. $\bar{c}c$ interpolators are not considered since we are interested in isospin $I=1$, while local diquark-antidiquark interpolators are also omitted as they seem to have very little influence, according to \cite{PRD_2015}. The interpolators used are of two types: charmonium-light meson, $H(|\mathbf{p}_i|^2)L(|\mathbf{p}_j|^2)$,
and $D$-meson-$D$-meson, $\bar{M}_i(|\mathbf{p}_i|^2)M_j(|\mathbf{p}_j|^2)$,
where every $H(\mathbf{p}_i)$, $L(\mathbf{p}_j)$, $\bar{M}_i(\mathbf{p}_i)$ and $M_j(\mathbf{p}_j)$ has an appropriate Dirac structure and is separately projected on to definite momentum $\mathbf{p}_i$, $\mathbf{p}_j$ so that the total momentum is $\mathbf{P}=\mathbf{p}_i+\mathbf{p}_j$. The full set of interpolating operators used are given in Tables \ref{tab:Cm} and \ref{tab:Cp}. They are constructed for $\Lambda^P=T_1^+$ and $\Lambda=A_2$, which are irreducible representations of the spatial lattice symmetry groups $O_h$ ($|\mathbf{P}|=0$) and $\textrm{Dic}_4$ ($|\mathbf{P}|=1\cdot2\pi/L$), respectively. The quantum numbers contributing to the chosen irreducible representations are not only $J^P=1^+$ but also unwanted higher $J=3$, \ldots and,  in the case of $\Lambda=A_2$, $J^P=0^-$, $2^-$. The Wick contractions are evaluated using the distillation method \cite{HadronSpectrum:2009krc} with 90 (100) Laplacian eigenvectors for $N_L=24$ (32).

\begin{table}[h!]
\centering
\scalebox{0.85}{
\begin{tabular}{|c|c|cc||cc|c|} 
\hline
\multicolumn{4}{|c||}{$|\mathbf{P}|^2=0,\ \Lambda^{PC}=T_1^{+-}$}             & \multicolumn{3}{c|}{$|\mathbf{P}|^2=1,\ \Lambda^{C}=A_2^{-}$}  \\ 
\hline
                      &                  & $J/\psi(0)\pi(0)$    & $\times 2$ & $J/\psi(1)\pi(0)$  & $\times 2$ &                              \\
                      &                  & $J/\psi(1)\pi(1)$    & $\times 2$ & $J/\psi(0)\pi(1)$  & $\times 2$ &                              \\
                      &                  & $J/\psi(2)\pi(2)$    & $\times 3$ & $J/\psi(2)\pi(1)$  & $\times 2$ &                              \\
$N_L=24$              &                  & $\eta_c(0)\rho(0)$   &            & $J/\psi(1)\pi(2)$  & $\times 2$ &                              \\
15 interpolators      & $N_L=32$         & $\eta_c(1)\rho(1)$   & $\times 2$ & $J/\psi(4)\pi(1)$  &            &                              \\
                      & 21 interp. & $\bar{D}^*(0)D(0)$   & $\times 2$ & $\eta_c(1)\rho(0)$ &            & $N_L=24$                     \\
                      &                  & $\bar{D}^*(1)D(1)$   & $\times 2$ & $\eta_c(0)\rho(1)$ &            & 21 interp.             \\
                      &                  & $\bar{D}^*(0)D^*(0)$ &            & $\eta_c(2)\rho(1)$ & $\times 2$ &                              \\
\cline{1-1}
\multicolumn{1}{c|}{} &                  & $J/\psi(3)\pi(3)$    & $\times 2$ & $\bar{D}^*(0)D(1)$ & $\times 2$ &                              \\ 
\multicolumn{1}{c|}{} &                  & $\eta_c(2)\rho(2)$   & $\times 3$ & $\bar{D}^*(1)D(0)$ & $\times 2$ &                              \\
\multicolumn{1}{c|}{} &                  & $h_c(1)\pi(1)$       &            & $\bar{D}^*(1)D(2)$ & $\times 2$ &                              \\
\multicolumn{1}{c|}{} &                  &                      &            & $\bar{D}^*(2)D(1)$ & $\times 2$ &                              \\
\cline{2-7}
\end{tabular}}
\caption{Table of interpolators transforming under irreducible representations $\Lambda^{PC}=T_1^{+-}$ and $\Lambda^{C}=A_2^{-}$ which correspond to $J^{PC}=1^{+-}$. All momenta here are in units of $2\pi/L$.}
\label{tab:Cm}
\end{table}

\begin{table}[h!]
\centering
\scalebox{0.85}{
\begin{tabular}{cccc||cc|c|c|} 
\hline
\multicolumn{4}{|c||}{$|\mathbf{P}|^2=0, \ \Lambda^{PC}=T_1^{++}$}                                                   & \multicolumn{4}{c|}{$|\mathbf{P}|^2=1, \ \Lambda^{C}=A_2^{+}$}                                  \\ 
\hline
\multicolumn{1}{|c|}{$N_L=24$}        & \multicolumn{1}{c|}{}                 & $J/\psi(0)\rho(0)$   &            & $\eta_c(1)a_0(0)$    &            &                  &                                        \\
\multicolumn{1}{|c|}{$\ 5\ $ interpolators} & \multicolumn{1}{c|}{}                 & $\bar{D}^*(0)D(0)$   & $\times 2$ & $\chi_{c0}(1)\pi(0)$ &            &                  &                                        \\
\multicolumn{1}{|c|}{}                & \multicolumn{1}{c|}{$N_L=32$}         & $\bar{D}^*(1)D(1)$   & $\times 2$ & $\chi_{c0}(0)\pi(1)$ &            &                  &                                        \\ 
\cline{1-1}
\multirow{10}{*}{}                    & \multicolumn{1}{|c|}{10 interp.} & $J/\psi(1)\rho(1)$   & $\times 3$ & $J/\psi(1)\rho(0)$   &            &                  & $N_L=24$                               \\
                                      & \multicolumn{1}{|c|}{}                 & $\chi_{c0}(1)\pi(1)$ &            & $J/\psi(0)\rho(1)$   &            &                  & 13 interp.                       \\
                                      & \multicolumn{1}{|c|}{}                 & $\chi_{c1}(1)\pi(1)$ &            & $\bar{D}^*(0)D(1)$   & $\times 2$ & $N_L=32$         &                                        \\ 
\cline{2-4}
                                      & \multirow{7}{*}{}                     &                      &            & $\bar{D}^*(1)D(0)$   & $\times 2$ & 17 interp.&                                        \\
                                      &                                       &                      &            & $\bar{D}^*(1)D(2)$   & $\times 2$ &                  &                                        \\
                                      &                                       &                      &            & $\bar{D}^*(2)D(1)$   & $\times 2$ &                  &                                        \\ 
\cline{8-8}
                                      &                                       &                      &            & $\eta_c(0)a_0(1)$    &            &                  & \multicolumn{1}{c}{\multirow{4}{*}{}}  \\
                                      &                                       &                      &            & $\chi_{c0}(2)\pi(1)$ &            &                  & \multicolumn{1}{c}{}                   \\
                                      &                                       &                      &            & $\chi_{c0}(4)\pi(1)$ &            &                  & \multicolumn{1}{c}{}                   \\
                                      &                                       &                      &            & $\chi_{c1}(2)\pi(1)$ &            &                  & \multicolumn{1}{c}{}                   \\
\cline{5-7}
\end{tabular}}
\caption{Table of interpolators transforming under irreducible representations $\Lambda^{PC}=T_1^{++}$ and $\Lambda^{C}=A_2^{+}$ which correspond to $J^{PC}=1^{++}$. All momenta here are in units of $2\pi/L$.}
\label{tab:Cp}
\end{table}

\section{Preliminary results}\label{sec:results}

\subsection{Energy levels}

We extract energy levels $E_n^{\textrm{lat}}$ from single-exponential fits to the eigenvalues $\lambda^{(n)}(t)\propto e^{-E_n^{\textrm{lat}}t}$ of the generalized eigenvalue problem \cite{Michael:1985ne}. They are shown in Fig. \ref{fig:1} for $1^{++}$ and Fig. \ref{fig:2} for $1^{+-}$. Many states lie below the lowest $D\bar{D}^*$ levels, in particular, for non-zero total momentum and the larger lattice volume. The energy shifts of states dominated by $D\bar{D}^*$ are very small in the $1^{++}$ case and negligible within the present uncertainties in the $1^{+-}$ case. Despite the light mesons $\rho$ and $a_0$ being resonances, they have been treated as stable particles. In the energy level plots in Figs. \ref{fig:1} and \ref{fig:2}, one can see that eigen-energies dominated by interpolators containing the $\rho$ meson have significant uncertainties.  

\subsection{$D\bar{D}^*$ scattering}\label{subsec:scattering}

\begin{figure*}[ht!]
\centering
\hspace{-6.5cm}
\begin{subfigure}{0.169\textwidth}
\includegraphics[width=1.3\textwidth]{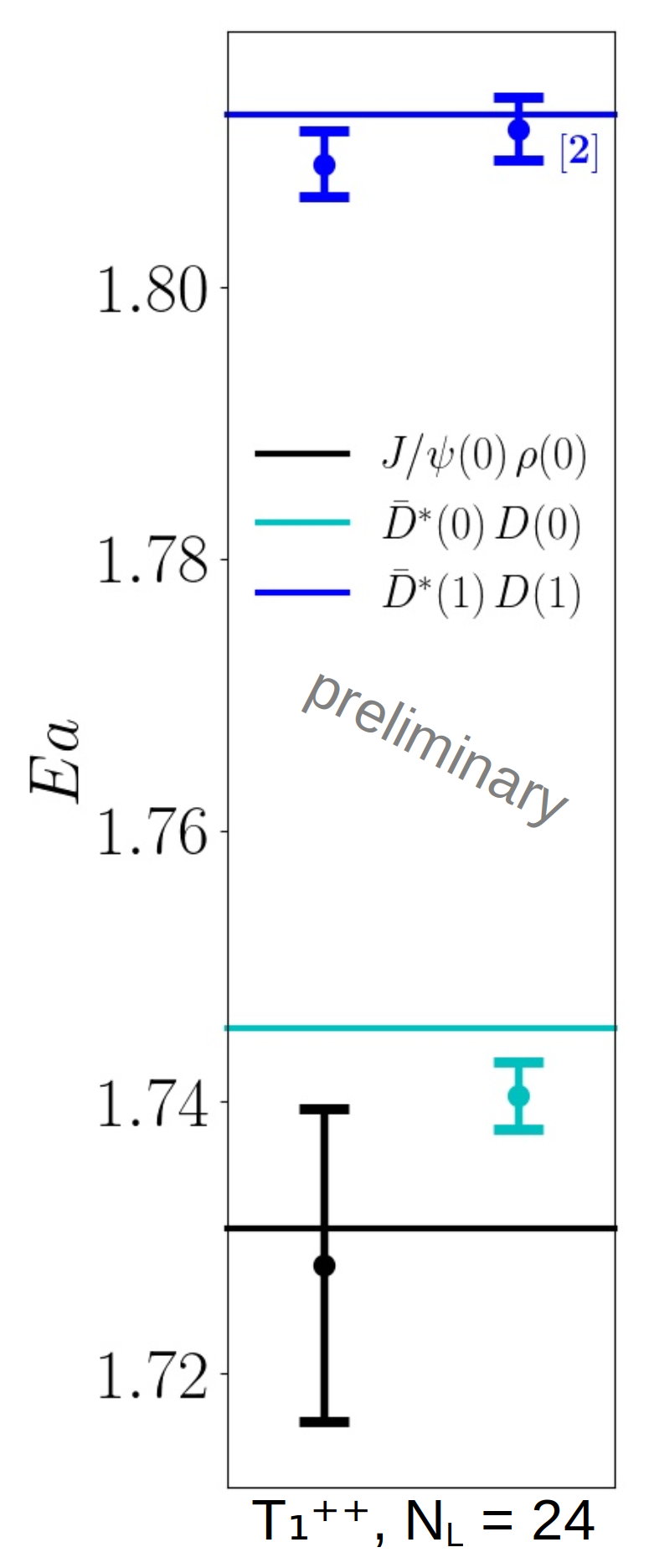}
\end{subfigure}
\hspace*{0.5cm}
\begin{subfigure}{0.159\textwidth}
\includegraphics[width=1.4\textwidth]{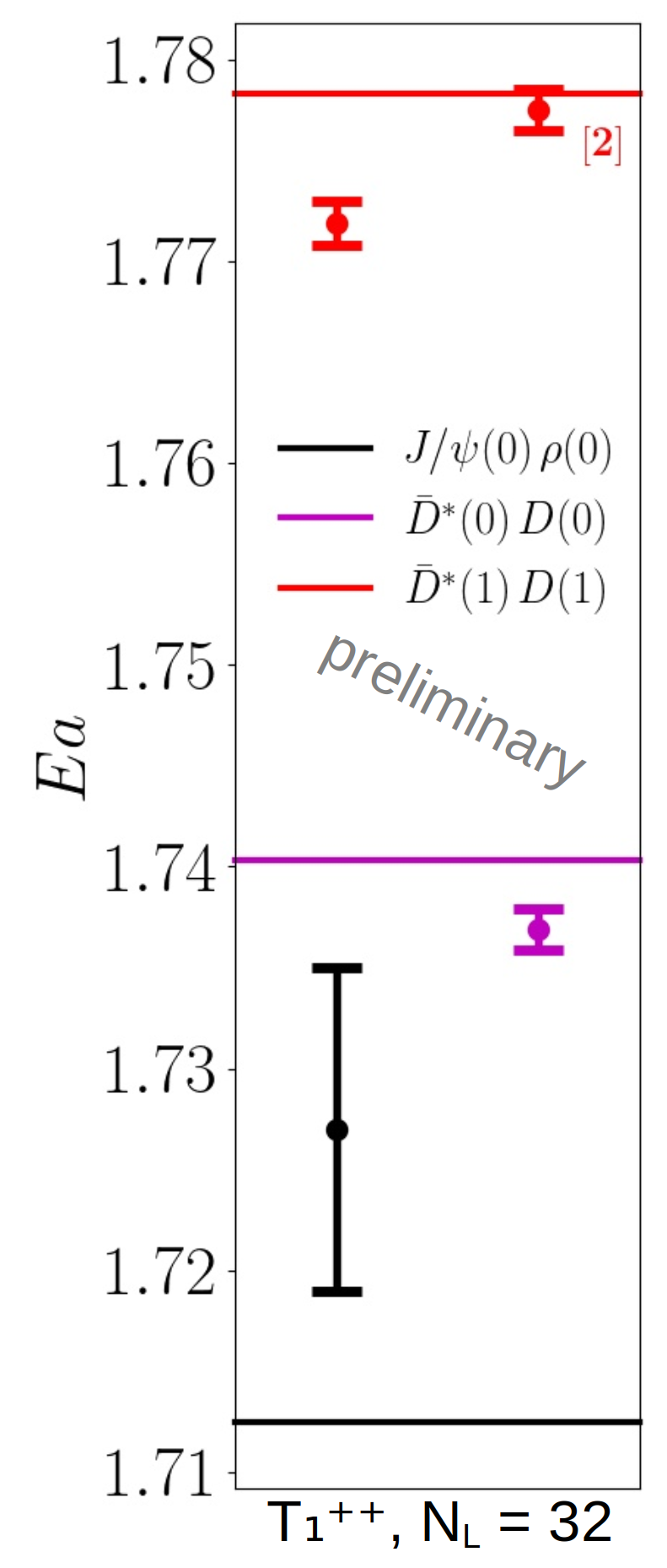}
\end{subfigure}
\hspace*{0.9cm}
\begin{subfigure}{0.29\textwidth}
\includegraphics[width=2.35\textwidth]{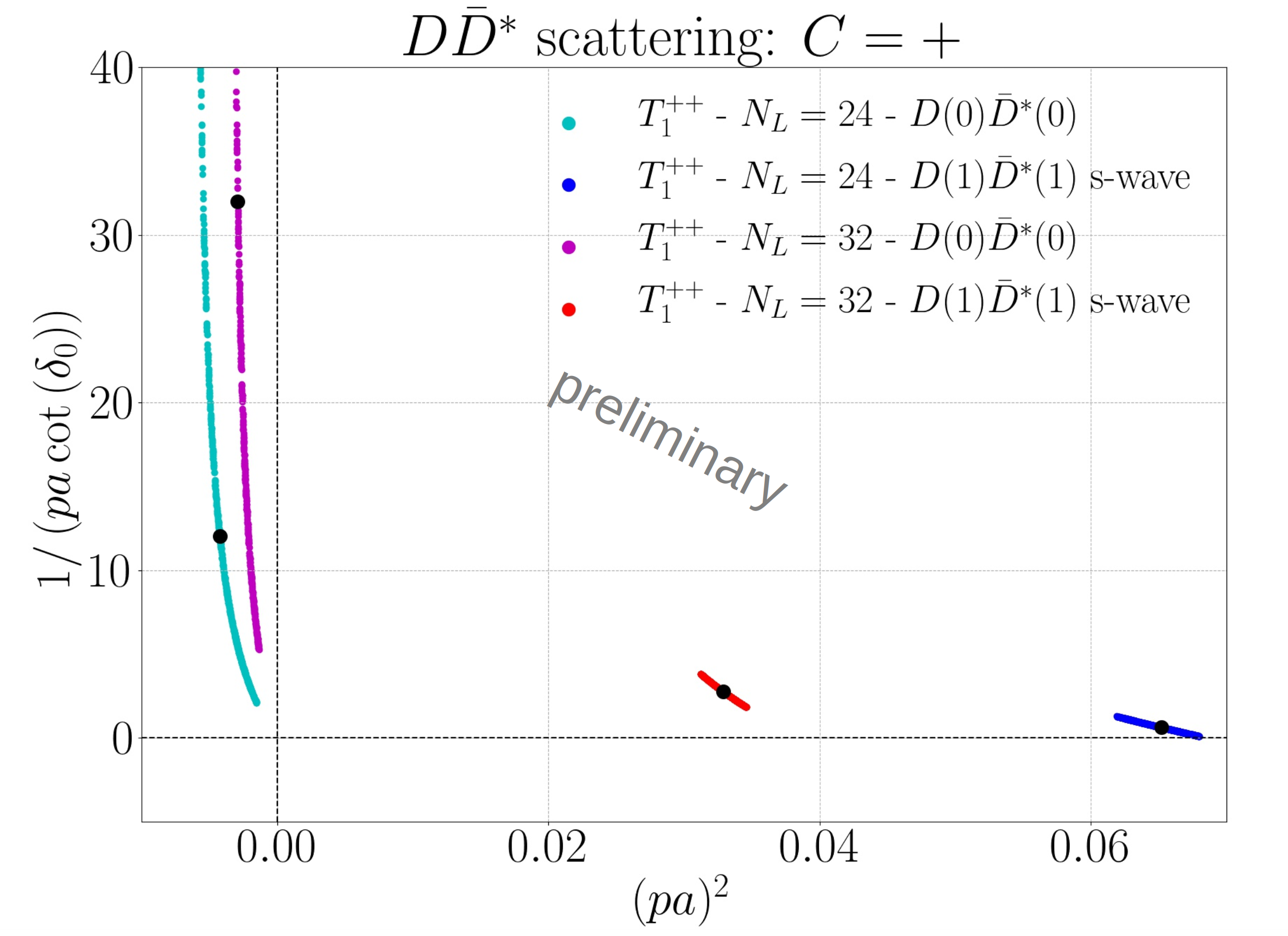}
\end{subfigure}
\caption{Results for $I(J^{PC})=1(1^{++})$. The two panes on the left represent the energy levels (points, $E_n^{\textrm{lat}}$) and non-interacting energies (lines, $E_{H_i(\mathbf{p}_i)}^{\textrm{lat}}+E_{H_j(\mathbf{p}_j)}^{\textrm{lat}}$). From left to right the panes represent $T_1^{++}$ with $N_L=24$ and $T_1^{++}$ with $N_L=32$, respectively. Numbers within the square brackets refer to the multiplicity of certain non-interacting levels. The plot on the right shows $1/(p\cot{(\delta}))$ where the colors of the states match those in the spectra, and $\delta$ is the s-wave $D\bar{D}^*$ scattering phase shift with approximations stated in subsection \ref{subsec:scattering}. Results are shown with $1\sigma$ statistical uncertainty.}
\label{fig:1}
\end{figure*}

\begin{figure*}[ht!]
\centering
\hspace{-6.cm}
\begin{subfigure}{0.145\textwidth}
\includegraphics[width=1.28\textwidth]{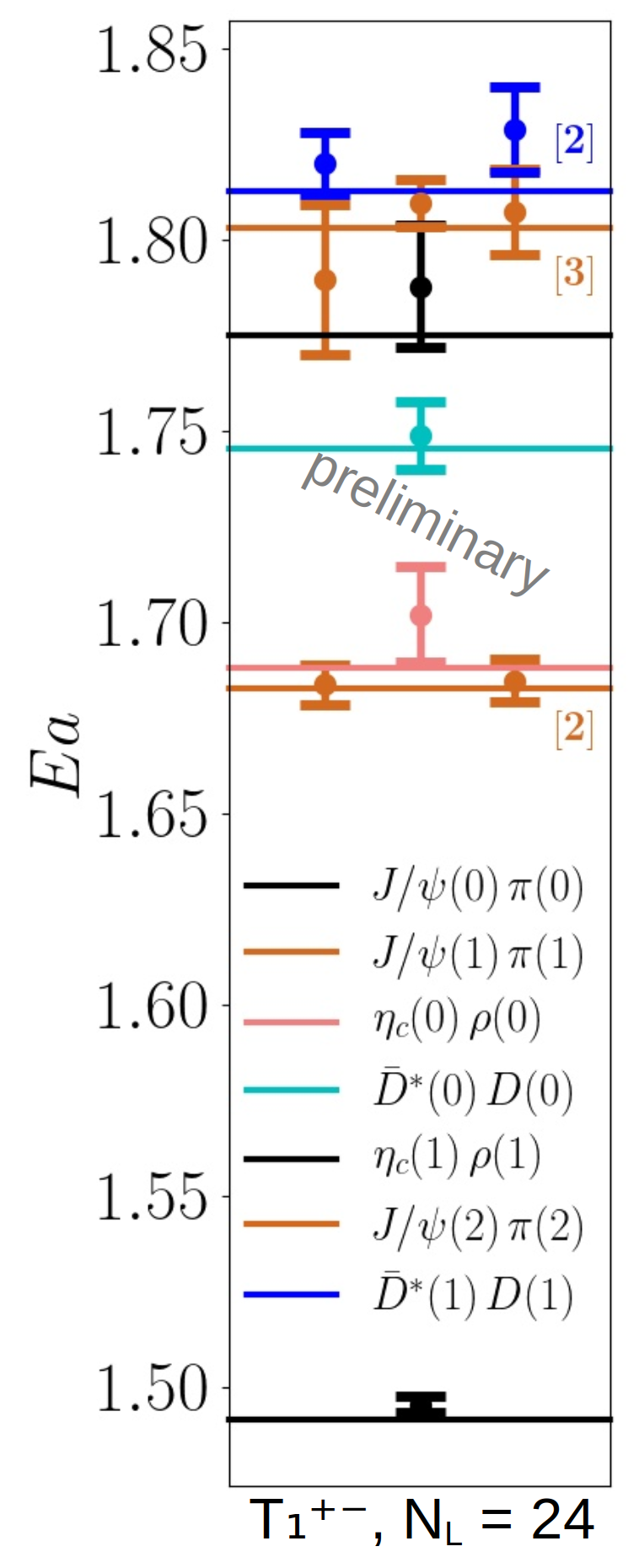}
\end{subfigure}
\hspace*{0.38cm}
\begin{subfigure}{0.141\textwidth}
\includegraphics[width=1.35\textwidth]{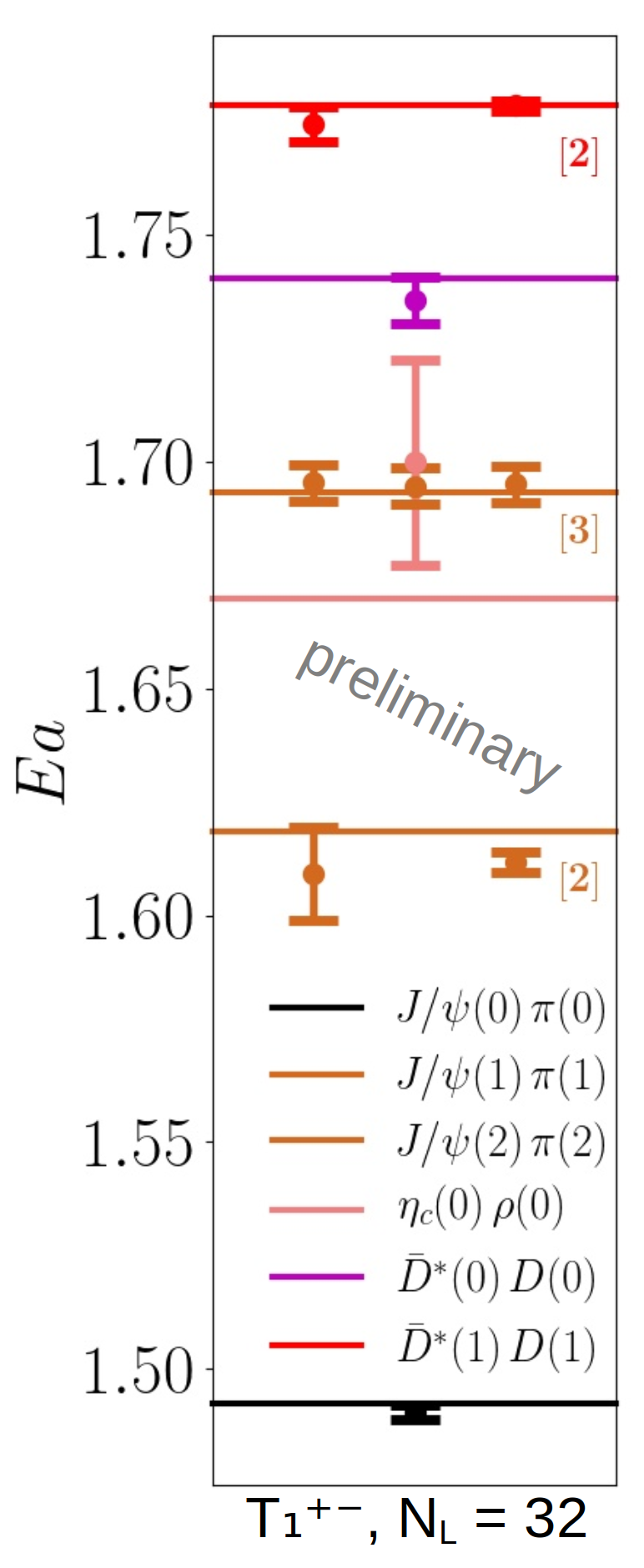}
\end{subfigure}
\hspace*{0.5cm}
\begin{subfigure}{0.148\textwidth}
\includegraphics[width=1.26\textwidth]{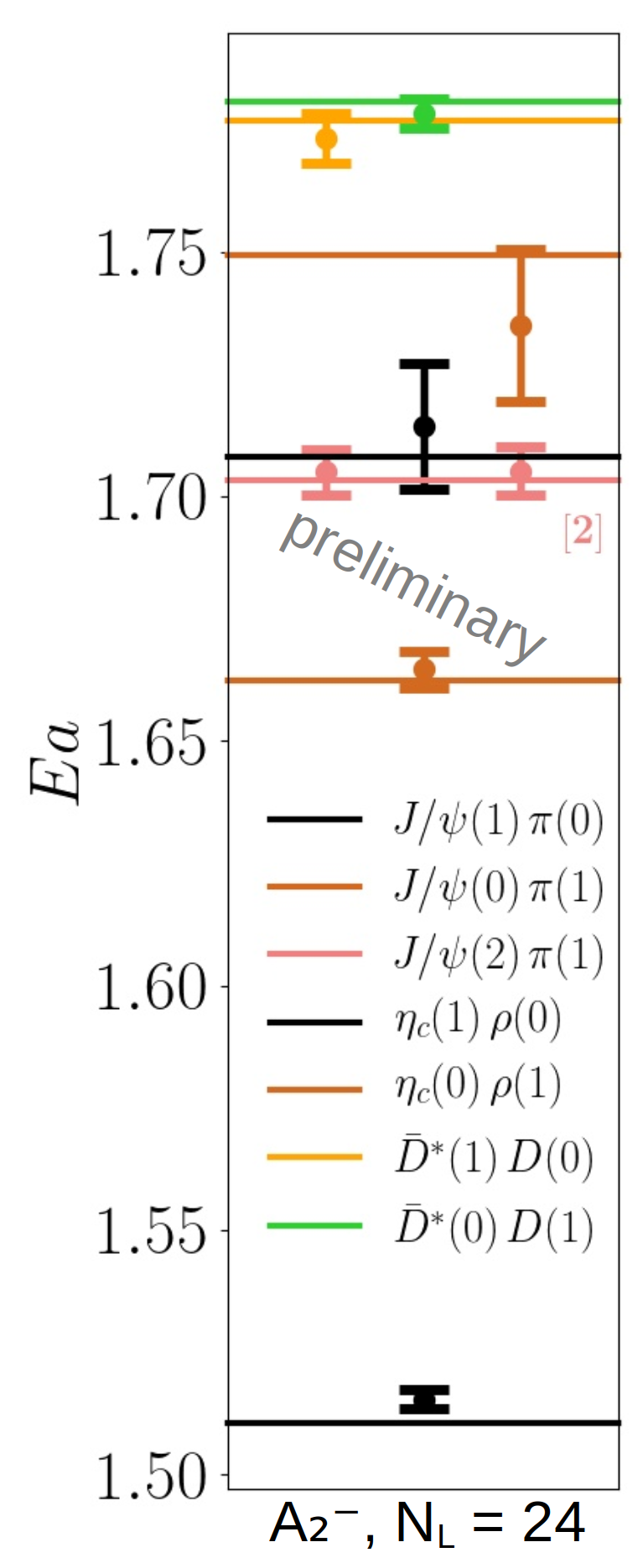}
\end{subfigure}
\hspace*{0.5cm}
\begin{subfigure}{0.28\textwidth}
\includegraphics[width=2.2\textwidth]{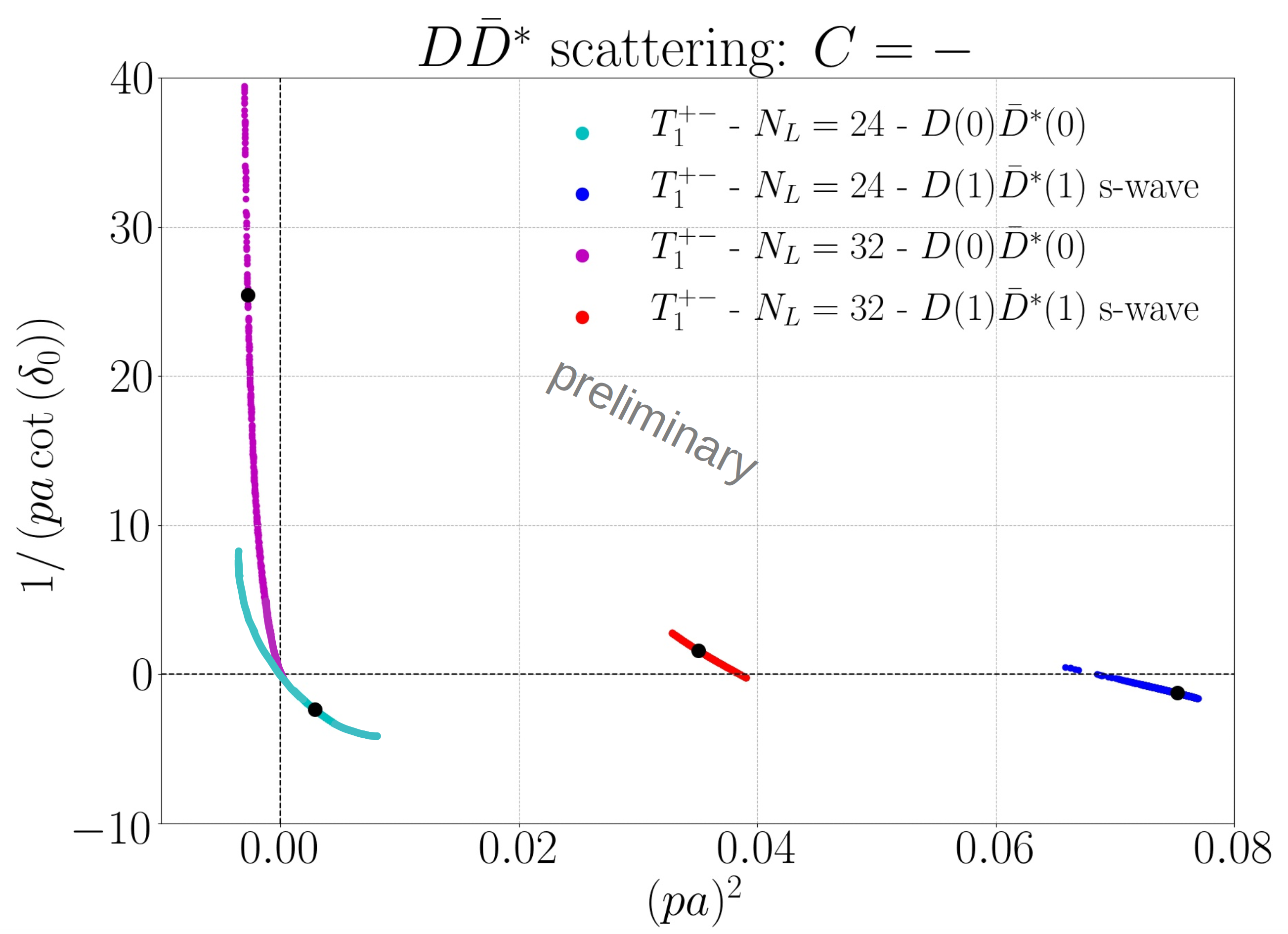}
\end{subfigure}
\caption{Results for $I(J^{PC})=1(1^{+-})$. The three panes on the left represent the energy levels (points, $E_n^{\textrm{lat}}$) and non-interacting energies (lines, $E_{H_i(\mathbf{p}_i)}^{\textrm{lat}}+E_{H_j(\mathbf{p}_j)}^{\textrm{lat}}$). From left to right the panes represent $T_1^{+-}$ with $N_L=24$, $T_1^{+-}$ with $N_L=32$ and $A_2^{-}$ with $N_L=24$, respectively. Numbers within the square brackets refer to the multiplicity of certain non-interacting levels. The plot on the right shows $1/(p\cot{(\delta}))$ where the colors of the states match those in the spectra, and $\delta$ is the s-wave $D\bar{D}^*$ scattering phase shift with approximations stated in subsection \ref{subsec:scattering}. Results are shown with $1\sigma$ statistical uncertainty.}
\label{fig:2}
\end{figure*}

To simplify the procedure, we focus on $D\bar{D}^*$ scattering near the threshold. Its coupling to other channels ($J/\psi\pi$, $\eta_c\rho$, $J/\psi\rho$) is neglected when studying the scattering amplitudes. The spectrum is expected to be dominated by the $\ell=0$ partial wave. The finite volume eigen-energies are connected to the infinite volume s-wave $D\bar{D}^*$ scattering phase shift $\delta$ via
\begin{equation}
p \cot{(\delta(p))}=\frac{2\mathcal{Z}^{\mathbf{d}}_{00}(1,(\frac{pL}{2\pi})^2)}{\gamma\sqrt{\pi}L}\>,
\label{E2}
\end{equation}
where higher partial waves are omitted, and the momentum $p=|\mathbf{p}_{cm}|$ in the center-of-mass frame is derived from
\begin{equation}
E_{\textrm{cm}}=\sqrt{|\mathbf{p}_{cm}|^2+m_i^2}+\sqrt{|\mathbf{p}_{cm}|^2+m_j^2}\>, \quad\textrm{where}\quad E_{\textrm{cm}}=\sqrt{E_n^2-|\mathbf{P}|^2}\>.
\label{E3}
\end{equation}
Discretization effects modify the dispersion relation, which deviates from the continuum one. To mitigate this, we use the following energies
\begin{equation}
E_n=E_n^{\textrm{lat}}+E_{H_i(\mathbf{p}_i)}^{\textrm{con}}+E_{H_j(\mathbf{p}_j)}^{\textrm{con}}-E_{H_i(\mathbf{p}_i)}^{\textrm{lat}}-E_{H_j(\mathbf{p}_j)}^{\textrm{lat}}\>,
\label{E4}
\end{equation}
where $E_{H(\mathbf{p})}^{\textrm{lat}}$ and $E_{H(\mathbf{p})}^{\textrm{con}}=(|\mathbf{p}|^2+m_H^2)^{1/2}$ are single-hadron energies. Within the aforementioned approximations, the scattering amplitude can be parametrized  in terms of $\delta$
\begin{equation}
T=\frac{1}{p \cot{(\delta(p))}-ip}\>.
\label{E5}
\end{equation}
Assuming elastic scattering near the threshold, one can perform the effective range expansion $p \cot{(\delta(p))}=1/a_0+r_0p^2/2+\mathcal{O}(p^4)$. Our preliminary results are presented in Figs. \ref{fig:1} and \ref{fig:2}. One can infer the smallness of the interaction from the small $1/(p\cot{(\delta}))$ values, which are zero in the non-interacting limit.

\section{Conclusion and outlook}\label{sec:conclusions}

We have extracted the spectrum of charmonium-like states with $1(1^{+})$. This is the first study considering hadronic states with these quantum numbers, a non-zero total momentum and two different lattice volumes. The energy shifts are small, which is consistent with conclusions from previous lattice QCD studies using the Lüscher method. This disfavors a significant attraction between $D$ and $\bar{D}^*$. Experimental evidence and findings from this preliminary study perhaps suggest that a significant coupling between channels causes the existence of $Z_c$. In the near future, we will make a comparison with phenomenological approaches and put constraints on them. In particular, we aim to compare our lattice eigen-energies with the energy levels that different models, such as \cite{Ortega2019,PhysRevD.90.016003}, predict. 

\acknowledgments
We thank David R. Entem, Feng-Kun Guo, Christoph Hanhart, Mikhail Mikhasenko, Daniel Mohler, Raquel Molina, Alexey V. Nefediev, Pablo G. Ortega, and Eulogio Oset for valuable discussions. M. S. acknowledges the financial support by Slovenian Research Agency ARRS (Grant No. 53647). S. C. acknowledges the support from the DFG grant SFB/TRR 55. The work of S. P. is supported by Slovenian Research Agency ARRS (research core funding No. P1-0035 and No. J1-8137) and (at the begining of the project) DFG grant No. SFB/TRR 55.

\bibliographystyle{JHEP}
\bibliography{bibliography_all_for_PoS}

\end{document}